\documentclass[global, twocolumn]{svjour}
\usepackage{draftcopy}
\usepackage[logonly]{trace}
\usepackage{verbatim}  % Needed for the "comment" environment to make LaTeX comments
\usepackage{vector}  % Allows "\bvec{}" and "\buvec{}" for "blackboard" style bold vectors in maths
\usepackage{textcomp}
\usepackage{amsmath}
\usepackage{amssymb}
\usepackage{amsfonts}
\usepackage{mathrsfs}
\usepackage{epstopdf}
\usepackage{graphicx}
\usepackage{color}

\def\Ket#1{\left|#1\right>}

\begin{document}

\title{EIT-Based Quantum Memory for Single Photons from Cavity-QED}

\author{M. Himsworth \and P. Nisbet \and J. Dilley \and G. Langfahl-Klabes and A. Kuhn}

\institute{Clarendon Laboratory, University of Oxford, UK.}
%\email{m.himsworth1@physics.ox.ac.uk}

\date{\today}

\maketitle

\begin{abstract}
We investigate the feasibility of implementing an elementary building block for quantum information processing. 
The combination of a deterministic single photon source based on vacuum stimulated adiabatic rapid passage, and a quantum memory based on electromagnetically induced transparency in atomic vapour is outlined. Both systems are able to produce and process temporally shaped wavepackets which provides a way to maintain the indistinguishability of retrieved and original photons. We also propose an efficient and robust `repeat-until-success' quantum computation scheme based on this hybrid architecture. 
\end{abstract}

%-----------------------------------------------------------------------------------------------------------------------------------------------

\section{Introduction}
To realise a practical quantum network, the quantum information carriers -- `flying qubits' -- between processing nodes must be deterministic, controllable, and resilient to decoherence. Linear-optics quantum computing (LOQC) provides such a route using measurement-based quantum computing with photons acting as qubits and passive optical elements for gate operations \cite{Knill2001,Kok2007}. The system is readily scalable with additional photon number states and can be incorporated into quantum key distribution networks. However, direct photon-photon interaction is weak and some quantum information processing (QIP) components, such as memory, are difficult to implement with passive elements alone, resulting in the need for extensive error handling \cite{Gingrich2003,Pittman2002,Ralph2005}. A more pragmatic and natural route to QIP is to use both atoms \emph{and} photons due to their stronger mutual interaction and suitability for different tasks. Within this hybrid architecture photons are used for LOQC-type processes and communication whereas atoms provide long stable storage times, efficient entanglement and state detection \cite{Lim2005,Duan2004,Cirac1997}. We see this hybrid system as an ideal candidate for `repeat until success' QIP which has been theoretically shown to be robustly fault tolerant and efficiently scalable \cite{Lim2006}. In this paper we assess the feasibility of a specific implementation of two elements which would form the main parts of a repeated quantum network block, namely a single photon source and a quantum buffer/memory \cite{Kimble2008}. 
\begin{figure}[!t]
\centering
\includegraphics[width=8.5cm]{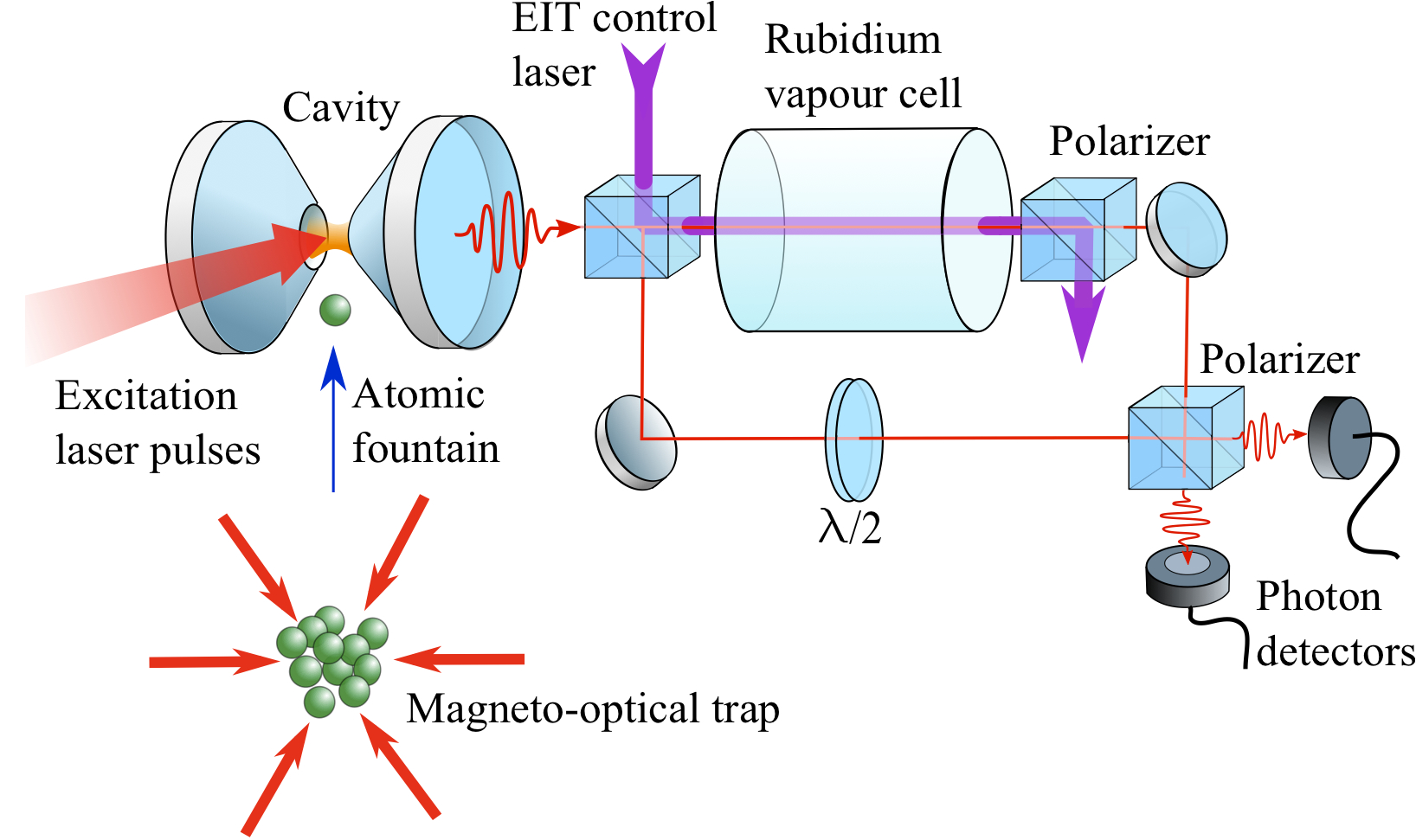}
\caption{Artist's view of a possible V-STIRAP single photon source and EIT storage arrangement. The atomic fountain is one method of interacting with unbound atoms (see text). Single photons emitted from a cavity are delayed or stored in a vapour cell. Once released, they may interfere with another photon from the cavity that travels directly to the beam splitter. Therefore two-photon interferences of the Hong-Ou-Mandel type will allow one to probe the coherence of the photon storage process.\label{CavityEIT}}
\end{figure}

A strong contender for a deterministic single photon source is vacuum stimulated Raman adiabatic passage (V-STIRAP) of a single atom in a high-Q cavity \cite{Kuhn2002}. In such a cavity-QED (C-QED) system, an atom with a $\Lambda$-type electronic structure undergoes an adiabatic Raman transition in which one coupling mode is a classical laser field and the other is the bare vacuum cavity mode. This constitutes a highly efficient deterministic photon source, also known as a `photon pistol'. Such cavity-confined atomic systems are also useful for photon storage and gate operations \cite{Duan2001,Raimond2001,Duan2005,Boozer2007,Maitre1997}.

Efficient and scalable QIP requires the flying qubits to be delayed or stored to synchronize gate operations and to allow extension to larger quantum networks. Passive elements such as cavities and delay lines offer a solution, but cannot be switched or adjusted quickly and suffer high loss over long delays \cite{Leung2006}. Coherent atomic states offer a popular solution in the form of electromagnetically induced transparency (EIT) \cite{Lukin2003} in which light pulses are stored as a collective dark superposition of ground states in a dense atomic ensemble or solid state impurities \cite{Kozhekin2000,Longdell2005,Yanik2004}. This method has also been shown experimentally to work in the single photon regime \cite{Duan2005,Chaneliere2005,Eisaman2005}. Other atomic ensemble quantum memory techniques are outlined in a recent review by Lvovsky \textit{et al}. \cite{Lvovsky2009}. 

The individual parts of the C-QED single photon source and EIT storage have seen much development, but as yet have not been combined to form a simple quantum network. A sketch of such a system is shown in Figure \ref{CavityEIT}. This paper will discuss the requirements for the individual elements of this system, what questions remain, how the system may be characterized, and how they can be implemented to form an advantageous path to QIP in the near future.

%---------------------------------------------------------------------------------------------------------------------------------
\section{Single Photon Source}

Quantum networks using photonic flying qubits require a source of single photons which can be emitted with a high probability into the QIP system on demand. These photons must have a predictable temporal shape to ensure time-reversibility for unitary gate operations, and maximal overlap of wavefunctions for entanglement. 
Many single photon sources are available: atomic cascades, parametric down conversion, weak laser beams, trapped ions, and semiconductor quantum dots, as described in recent reviews \cite{Kuhn2010,Lounis2005,Santori2010}. All of the above techniques demonstrate the sub-Poissonian statistics (anti-bunching) of a single photon source, but this does not necessarily mean that the photons are either deterministic, or indistinguishable. Techniques such as atomic cascade \cite{Clauser1978,Grangier1986}, EIT \cite{Eisaman2005} and parametric down conversion \cite{Tapster1988,Hong1985} provide `herald' photons which indicate the presence of another photon in a separate mode, but the initial photon emission is probabilistic and therefore of limited for QIP.

\begin{figure}[!t]
\centering
\includegraphics[width=8.5cm]{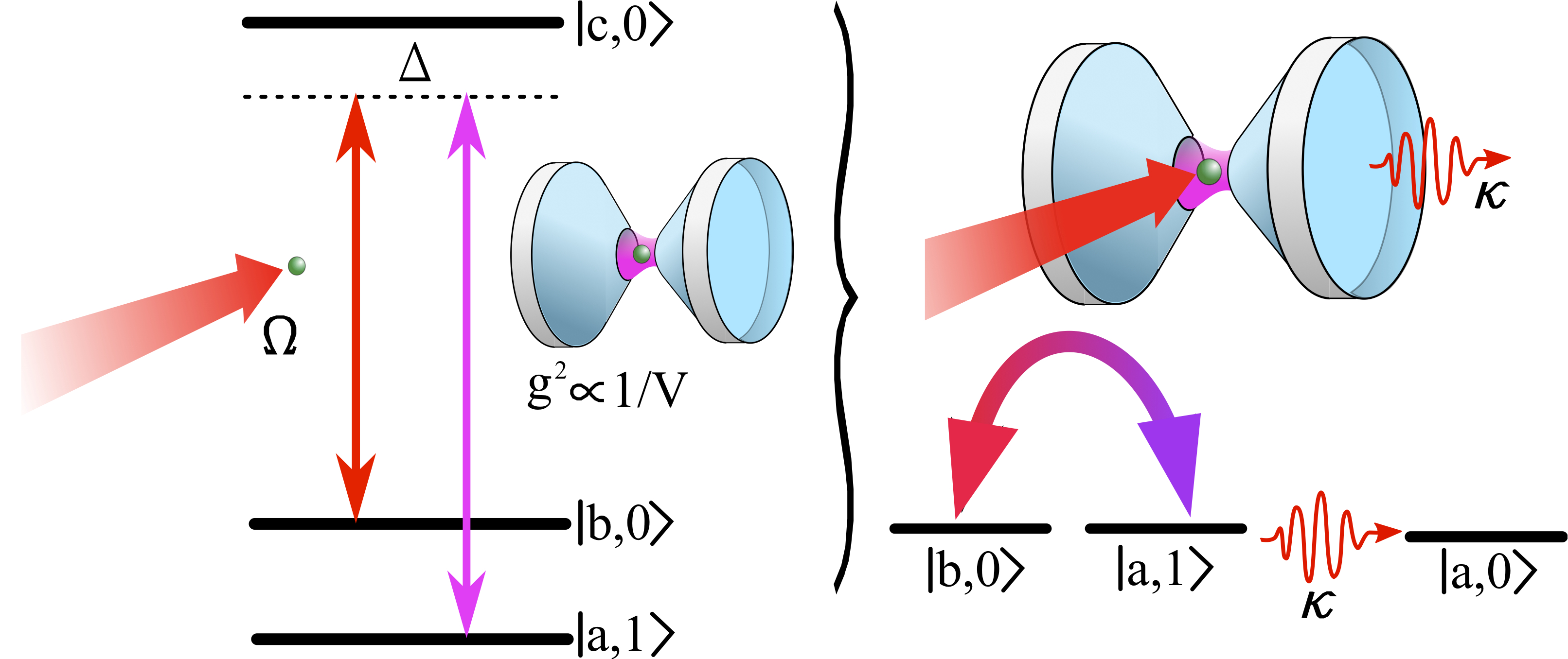}
\caption{Energy level scheme for V-STIRAP produced photons.  A laser beam $\Omega$ and the cavity vacuum mode $g$ provide the two arms of a Raman transition. The atom is adiabatically driven via a dark state which decouples the excited state $\Ket{c,0}$ and so avoids irreversible spontaneous processes. The Raman process is slow compared to the cavity decay $\kappa$ so that once the photon is produced it is emitted from the cavity.\label{Stirap}}   
\end{figure}

A single atom which is coupled to a high-finesse optical cavity and exposed to classical laser pulses can be used to emit or absorb single photons in a controlled manner \cite{Kuhn2002}. A photon is generated by an adiabatically driven Raman transition, with the vacuum field of the cavity stimulating one branch of the transition, and laser pulses driving the other (See Fig. \ref{Stirap}) \cite{Kuhn1999,Law1996}. The cavity enhances the probability of photons being emitted into a well-defined mode of the radiation field which is coupled to the outside world. Anti-bunching is found in the intensity correlation of the light, demonstrating that a single atom emits photons one-by-one. The photons have properties close to laser light: they propagate along one mode, they all have the same frequency and cannot be distinguished from one another. This particular feature makes these photons well suited for quantum information processing. Photons produced in this manner also have the unique property such that their temporal profile can be accurately tailored, and therefore optimized, for a specific role \cite{Vasilev2010}. For example, generating twin peaked `time-bin' pulses for robust propagation of information \cite{Brendel1999,Gisin2002}. Therefore, in this section, we shall discuss the design requirements for an ideal atom-cavity single photon source.

\begin{figure}[!t]
\centering
\includegraphics[width=10cm]{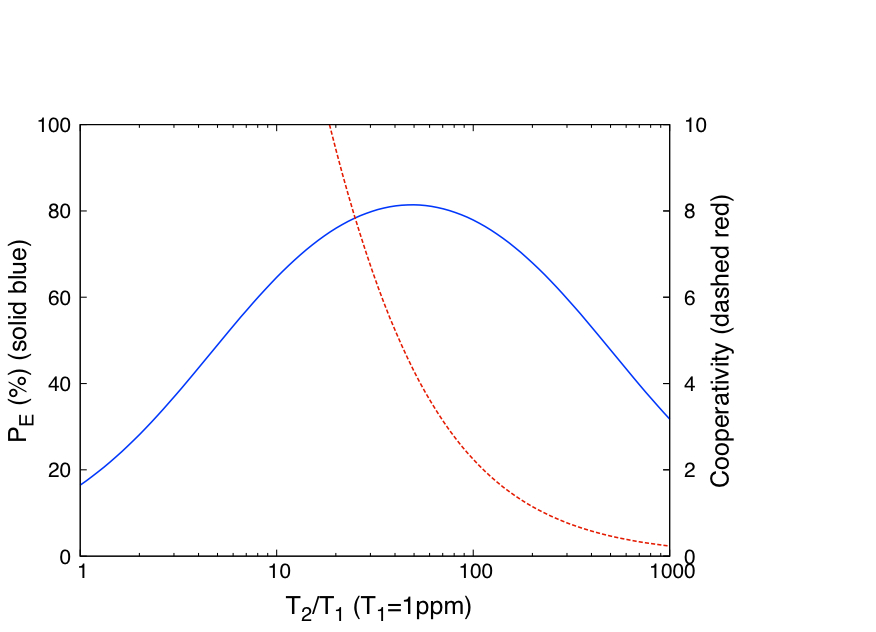}
\caption{The effect upon the photon emission probability (Eqn. \ref{Pemit}) and cooperativity (Eqn. \ref{coop}) in a Fabry-Perot cavity with asymmetric mirror transmittance, $T_{1/2}$. The plot shows that the cavity can have a high probability of emitting a photon whilst remaining in the strong coupling regime. The scattered losses are 2\,ppm per mirror and the diameter and length of the cavity mode is 20$\,\mu$m and $100\,\mu$m, respectively. $g,\gamma=\{15,3\}\,$MHz. \label{photonemit}}
\end{figure}

Using an atom-cavity system as a single photon source relies on two properties which are not mutually compatible. 
Firstly, the interaction must be unitary so there must be strong coupling $g>\{\kappa,\gamma\}$
where $g$ is the atom-cavity coupling strength, and $\kappa$  and $\gamma$ are the respective cavity and atomic decay rates. From standard definitions (for example see \cite{Walther2006}) this requires that the cavity mirrors should have the highest reflectivity possible and the cavity volume, $V$, should be small. The second property is that photons must be emitted from the system through the cavity mirrors with a high probability and a high out-coupling rate, hence a low finesse (or low reflectivity) cavity is required. These hurdles are routinely overcome by using an asymmetric cavity where one mirror has a significantly higher transmission than the other. The photon emission probability \cite{Law1997,Kuhn2010} using an asymmetric cavity is
\begin{equation}
P_{E}=\frac{T_{2}}{T_{1}+T_{2}+2H}\left(\frac{g^{2}}{\gamma\kappa+g^{2}}\right) \label{Pemit},
\end{equation}
where $T_{1/2}$ is the mirror transmittance and $H$ is the fraction of photons lost through scattering from each mirror. The term in parentheses describes the probability for a photon to be spontaneously emitted into the cavity decay mode, $\kappa$, and the prefactor scales the proportion of this decay through $T_{2}$, the outcoupling mirror. The cooperativity,
\begin{equation}
C=\frac{g^{2}}{2\kappa\gamma}\label{coop},
\end{equation}
signifies the atom-cavity coupling strength and must be greater than unity to be in the strong coupling regime. Figure \ref{photonemit} shows $P_{E}$ and $C$ plotted against reflectivity asymmetry for a realistic system; we see that there is a region where strong coupling and high emission probability occur simultaneously.

Several methods are employed to manoeuvre single atoms into the cavity mode: atoms may be dropped from above \cite{Kuhn2002}, propelled upward in an atomic fountain \cite{Munstermann1999}, confined within the cavity using a dipole trap \cite{McKeever2004} and/or transferred by an optical conveyor belt \cite{Nussmann2005}. A dipole trap offers the longest interaction times, as demonstrated by Hijlkema \textit{et al.} \cite{Hijlkema2007} who achieved a 30 second-long photon stream from a single atom. However, the AC-Stark shifts of the internal quantum states must be accounted for and the additional fields required to address the atom increase the complexity of the system. There are further complications in the form of thermal motion \cite{Maunz2004,Nussmann2005,Kubanek2009}, vibrational states and anti-trapping excited states, all of which make it hard to realise an undistorted 3-level system.

A natural choice would be to use trapped ions to ensure long interaction times and positioning with greater precision than optical dipole traps \cite{Mundt2002,Keller2004}. The ions are, however, sensitive to external fields and incompatible with confinement close to dielectric surfaces. So far this has limited the cavity mirror separation to large distances and therefore results in large mode volumes and weak couplings. Also, the laser frequencies required for interacting with ions will be different from that used in the EIT storage. Even so, ion-cavity photon sources have demonstrated impressive results \cite{Keller2004,Barros2009}.

To achieve long interaction times of `free' atoms in the cavity the atomic fountain method appears to be the most suitable choice. The atoms are initially cooled in a magneto optical trap (MOT) positioned below the cavity and launched upwards against gravity so that the turning point of their passage coincides with the cavity mode. Launching may be achieved by detuning the vertical MOT beams so that the atoms feel an upward force (the so-called `moving-molasses' technique \cite{Munstermann1999,Berthoud1999}), or by pulsing external magnetic fields \cite{Kim2002}. The time spent within the cavity mode by an atom dropped from above is on the order of a hundred microseconds, whereas it may be increased by an order of magnitude using the fountain method. The interaction time may be controlled by adjusting the launch velocity, and a continuous fountain of atoms could be produced from a 2D MOT thus reducing the dead time \cite{Berthoud1998}. We must concede that this loading method is probabilistic, but once an atom is found within the cavity mode the stream of photon emissions is deterministic during the interaction time. We discuss later how this probabilistic loading may not be a significant problem when coupled to an EIT memory for repeat-until success quantum computing.

\begin{figure}[!t]
\centering
\includegraphics[width=8.5cm]{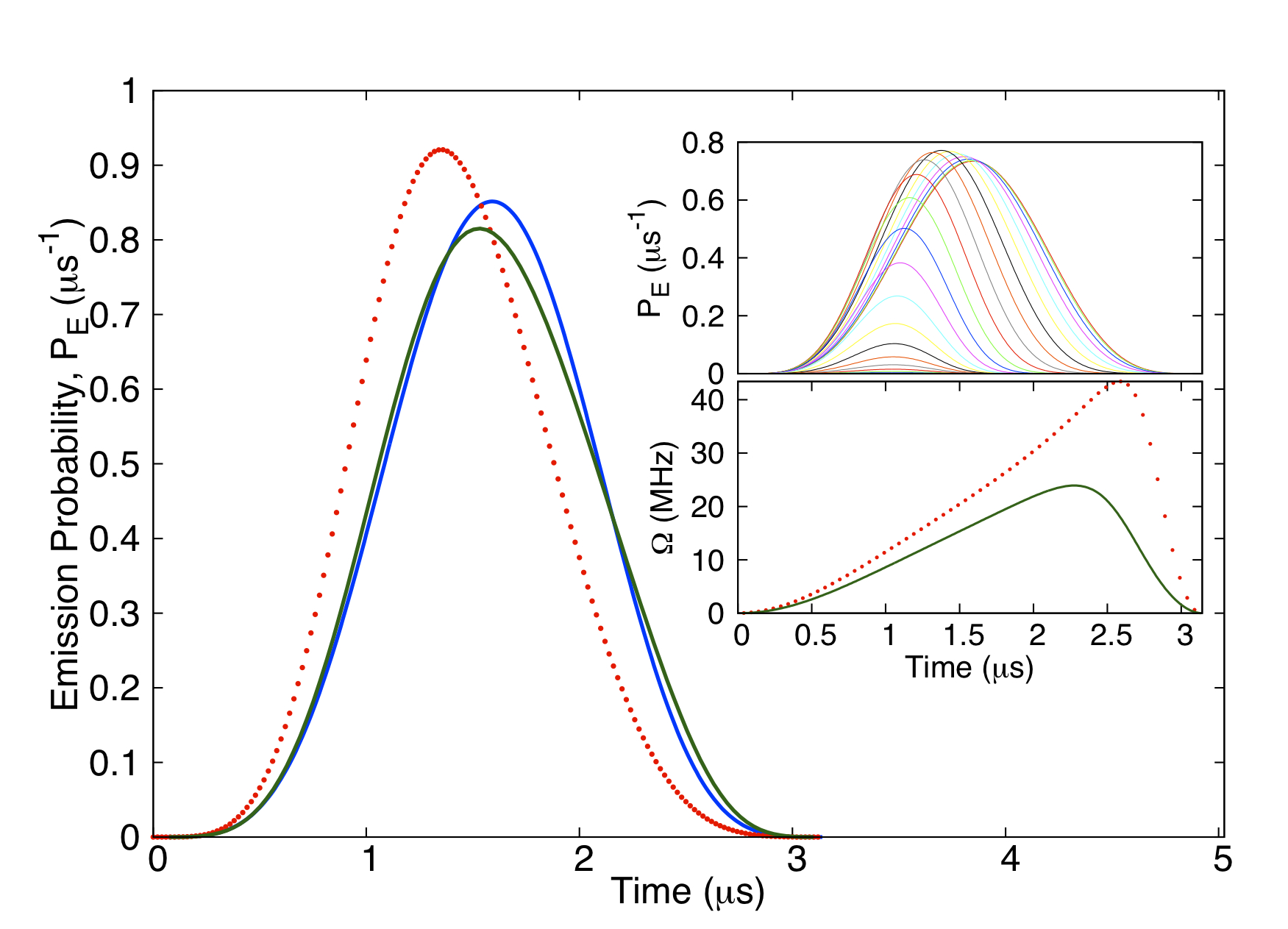}
\caption{The emission probability and temporal shape of the photon can be controlled by the ratio of the drive laser intensity, $\Omega(t)$, and the atom-cavity coupling, $g$. Launching 'free' (untrapped) atoms into the cavity mode will result in a distribution of $g$'s which will impact on photon interference experiments. The main plot shows the average photon shape (squared probability and normalised to unit area) calculated by the method in Ref. \cite{Vasilev2010} for: atoms situated only at the strongest coupling $g_{max}$ (thick solid blue line), atoms equally distributed across the mode (dotted red line), and atoms equally distributed across the mode but with $\Omega(t)$ optimised for $0.75\,g_{max}$ (thin solid green line).
For the distributed photons, the different values of $g$ have been binned and weighted by the fraction of the cavity mode volume that each bin occupies. We see that photons tailored for $g_{max}$ become distorted when averaging over atoms experiencing different $g$. Most significantly, the photon peak moves to earlier times due to the increasing ratio of $\Omega(t)/g$ (assuming $\Omega(t)$ is uniform across the cavity mode). As the cavity mode volume fraction is greater for weaker $g$; the photons start to regain the tailored shape when $\Omega(t)$ (lower inset) is calculated for a lower value of $g$. The upper inset shows the photon emission probability for different values of $g$. For these plots $g_{max},\kappa, \gamma=\{15,12,3\}\,$MHz. \label{shapes}}
\end{figure}

It is in the stochastic nature of the process that the atoms will not be reliably placed into the center of the cavity mode resulting in a variation of $g$, which in turn affects the photon emission rate. The photon emission probability $P_{E}$ (Eqn. \ref{Pemit}) predicts a decreased sensitivity of $P_{E}$ with $g$, but only in the very strong coupling regime. This may not be practical when high photon emission rates are required (as discussed earlier). The effect of the variation of $g$ on the photon shape and emission probability is shown in Fig. \ref{shapes} for a realistic system. We see that by using optimization techniques \cite{Vasilev2010} for the drive pulse, but not for the peak cavity coupling rate, it is possible to generate a statistically indistinguishable train of photons. One possible method to improve atom loading is to use the cavity stabilization light, blue-detuned in a higher `donut mode' to funnel atoms into the region of highest $g$ \cite{Nussmann2005}.

The single photon emission rate depends upon the dynamics of the V-STIRAP process, the cavity decay rate, and any repumping requirements to `reset' the atom. The latter timescale is normally the longest, lasting several microseconds, thus limiting the repetition rate to hundreds of kHz \cite{Hijlkema2007}. A technique to produce polarization-controlled photons from the cavity by shifting the laser frequency between Zeeman sub-states removes the need for repumping and can increase the repetition rate to MHz \cite{Wilk2007prl,Gogyan2010}. 

The temporal length of the photon, $\tau$, is defined by the rate of adiabatic passage between the ground states and is controlled by the Rabi frequencies of the cavity coupling $g$, and laser field, $\Omega$. In order to remain in the adiabatic condition (i.e. negligible steady state population in the excited state) and to achieve shorter photon pulse lengths, the laser field strength must be increased ($g$ is assumed constant and $\Omega_{max},g\gg \tau^{-1}$). The photon temporal length does not directly depend upon the cavity decay rate, however $\kappa$ should not be so low that the photon remains within the cavity thus increasing the possibility of re-absorption by the atom. The photon length is usually on the order of a microsecond in experimental demonstrations to date \cite{Kuhn2002}. 

From the above, it follows that the photon's temporal \textit{profile} can be controlled with the laser field \cite{Keller2004}. Theoretical analysis of this process has identified a method of reverse engineering the laser-atom Rabi frequency, $\Omega(t)$, to arbitrarily shape the output pulse to match a predetermined shape and efficiency with high accuracy \cite{Vasilev2010}. To the author's knowledge, as yet no experimental demonstrations of photon pulse shaping by this method have been carried out, although it has been shown to accurately fit the numerical model. A directly related situation is the ability of the cavity system to absorb a photon, which is vital for QIP \cite{Kimble2008}. For instance, the time reversal of the photon emission process would allow reabsorption of the photon, something which has been shown theoretically for time-symmetrical photon wavepackets \cite{Cirac1997,Fleischhauer2000}.

The feature which has so far prevented atom-cavity systems from demonstrating a scalable network is the complexity and size of the apparatus involved. Advances in tunable microfabricated cavities for atom detection \cite{Trupke2005,Aoki2006} and cavities formed from optical fibers \cite{Steinmetz2006} may provide a solution. 

%---------------------------------------------------------------------------------------------------------------------------------
\section{Quantum Memory}

An essential feature for quantum networking is the mapping of single photon states onto long-lived atomic quantum states and vice versa: to synchronize various components, and to extend the network over large distances using quantum repeaters \cite{Dur1999}. A promising scheme is to store the photon state onto a single trapped atom, however the resilience of the trap to perturbations and decoherence cannot be guaranteed \cite{Boozer2007,Simon2010}. A more robust approach is to make a superposition of the photon throughout an ensemble of atoms in which the loss of a single atom has a negligible effect on the photon retrieval. A common route to achieving this is electromagnetically induced transparency in which a strong `control' laser coherently affects the transmission of a weak `signal' beam through a medium \cite{Fleischhauer2005}. Atoms with $\Lambda$-type electronic structures can be coherently excited into a superposition of ground states which are immune to spontaneous decay. In the terminology of quasi-particles these are called dark state polaritons or spin waves \cite{Fleischhauer2002}. 

In a spectral scan of the weak signal beam across a resonance a narrow EIT peak is found, whose width predominantly depends on the control laser strength and can lead to complete transparency at resonance. Accordingly, this effect also leads to a sharp variation of refractive index across resonance which can strongly affect the group velocity of the signal pulse according to
\begin{equation}
v_{g}=\frac{\partial \omega}{\partial k}\simeq \frac{2|\Omega_{c}|^{2}}{\alpha\gamma_{P}}, 
\end{equation} 
where $\Omega_{c}$ is the Rabi frequency of the control field, $\alpha$ is the absorption coefficient \cite{Phillips2008} and $\gamma_{P}$ is the (single photon) depolarization rate. Therefore it is possible to slow the pulse down and store it entirely within the medium. 

Several groups have demonstrated the storage and retrieval of light pulses -- down to the single photon level -- in several mediums, such as thermal vapours, ultra cold atomic clouds and solid state impurities \cite{Simon2010}. The limit of the storage period depends upon the coherence lifetime of the spin wave which can be hundreds of milliseconds in atomic vapours \cite{Julsgaard2004} and up to seconds \cite{Longdell2005} in cryogenically cooled solid state impurities \cite{Longdell2005}. Therefore EIT is seen by many as a promising candidate for a quantum memory or buffer \cite{Lvovsky2009,Simon2010}. 

In order to retain compatibility with the atom-cavity single photon source the obvious choice is to use the same atomic species for storage, namely alkali atoms. Ultracold atoms offer long coherence lifetimes, high optical depth, negligible Doppler broadening and have demonstrated single photon storage \cite{Chaneliere2005},  however the technical complexity of MOTs makes them an unsuitable candidate for scalable memory devices; also the inhomogeneous magnetic fields can lead to significant decoherence and the physical size of the atom cloud is limited. Storage in thermal vapour, on the other hand, has probably seen the most development, it is comparably simpler to implement, and the effects of Doppler broadening can be mitigated. 

Before we look at the specific implementation of light storage in a thermal vapour we shall briefly discuss the general technicalities of using EIT as a quantum memory. 
The most important question to consider in this system is how efficiently a photon emitted from the cavity can be stored within a medium of length $L$, and whether the photon's spectrum will fit within the EIT transparency window. 
A light pulse with a temporal width, $\tau$, will be spatially compressed by a factor $c/v_{g}$ as it enters the EIT medium. The control beam intensity is inversely proportional to the group velocity, and proportional to the transparency spectral window. Hence there exists boundaries to the photon's shape (assuming the photon vacuum bandwidth$\,=\,1/\tau$)
\begin{equation}
\frac{v_{g}}{L}\ll \frac{1}{\tau}\ll v_{g}\sqrt{\frac{\alpha}{L}}\label{bounds},
\end{equation}
where the lower bound is the vapour cell length and the upper bound the transparency window. We see that both limitations fundamentally depend upon the optical depth, $d=\alpha L$, which must be much greater than unity for efficient storage \cite{Fleischhauer2002,Gorshkov2007a}. 
Once the photon is completely located within the vapour cell the control field is adiabatically ramped to zero and the photon state is mapped onto the atomic dark state. The adiabaticity condition for the timescale on which to map the photon onto the spin wave is \cite{Gorshkov2007a}
\begin{equation}
\tau d \gamma_{P}\gg 1.
\end{equation}
The spin wave may be mapped back into a photon by reversing the writing procedure with the resulting direction of propagation relying on the wavevector of the control beam \cite{Zibrov2002}. Hence it is possible to retrieve the photon back along its original path. This property can be used as another method to mitigate the non-uniform pulses emitted from the cavity: by reversing the photon back into the cavity, one can ensure efficient coupling back to the (same) atom by reversing the V-STIRAP laser intensity modulation. This `phase conjugate mirror' would provide an efficient process for gate operations without the need for knowledge of the exact atom-photon coupling efficiencies or photon shapes. 

The efficiency of the storage, $\eta$, is calculated by the ratio of the output and input energy of the signal pulses. Fundamentally, 100\% storage and retrieval is possible \cite{Fleischhauer2002}, however in practice at large optical depths ($d>25$) the efficiency is reduced due to competing processes such as four wave mixing, spin exchange collisions and radiation trapping \cite{Phillips2008}. As discussed earlier, the EIT storage process requires a trade-off between a strong control beam to reduce absorption and a weak control beam in order to slow the pulse so that it may fit inside the cell. Therefore, avoiding leakage and/or absorption of the signal pulse at practical optical depths is difficult.  With simple control beam dynamics (such as a step function) a significant proportion of the the photon is mapped onto the atoms in the first section of the medium and subsequent restoration back into a optical pulse requires the photon to pass through the remaining length of the medium which can result in an additional loss of the signal. Therefore, retrieval (or reading) of a stored photon in the opposite direction of the writing stage is preferred over copropagating read and write control fields \cite{Gorshkov2007a}. Studies have shown that there exists an optimum distribution of the spin wave across the medium for a specific optical depth and photon shape. The optimal spin wave can be achieved by tailoring the dynamics of the control beam intensity \cite{Novikova2008} in much the same manner as the single photon shaping from a cavity. For optical depths in the region $d=10-20$, the maximum storage and retrieval efficiency, $\eta$, is found to be $65\%$ in the co-propagating read/write beam direction and $80\%$  in the counter-propagating direction \cite{Gorshkov2007b}. 

The effect of spin wave decay, by diffusion of atoms out of the beams or dephasing by non-uniform magnetic fields, on the detected storage efficiency follows an exponential dependence, $\eta \exp(-2\gamma_{S}T)$, where $T$ is the storage time and $\gamma_{S}$ is the spin wave decay rate \cite{Gorshkov2007a,Mewes2005}. In a thermal vapour, the limiting timescale for spin wave coherence is defined by the diffusion of the atoms away from the beam and ultimately colliding and dephasing with the cell walls \cite{Figueroa2006}. Buffer gas is routinely added to the cells to reduce the mean free path of the alkali atoms by polarization maintaining collisions \cite{Firstenberg2008}, and the walls of the cell may be coated in a thin film (e.g. paraffin) to allow atoms to elastically collide several thousand times without suffering significant decoherence \cite{Wang2010}. The dark state polariton is shielded from the dephasing effects of collisions because it is mapped on a nuclear moment which has a negligible collisional cross-section compared to the electronic moments. Only at very high buffer gas pressures (greater than 100 Torr) pressure broadening will result in lower optical depths and the very fast dephasing disrupting the read/write processes \cite{Manz2007}. 

Although  the use of co-propagating signal and control beams during the writing stage can result in negligible Doppler shifts, there maybe residual Doppler sensitivity due to the choice of the ground states used for storage. The ground state hyperfine splitting of alkali atoms is on the order of GHz which result in the difference in wavevectors, $\Delta k=k_{c}-k_{s}$ between the control, $k_{c}$, and signal beams, $k_{s}$, on the order of centimeters \cite{Klein2009}. Any atom traveling beyond this distance during storage will be out of phase during the read stage. Again the addition of buffer gas at typical pressures used in EIT (tens of Torr) results in a diffusion dominated homogeneous broadening and thus the residual Doppler effects become negligible \cite{Klein2009,Gorshkov2007b}.  Moreover, the presence of a buffer gas reduces the sensitivity to the alignment between control and signal beams (and thus wavevector mismatch) and can facilitate separation of the weak single photon signal beam from the strong control beam. Storing photons between Zeeman-shifted substates also offer negligible residual Doppler shifts,  but one must also note that the existence of nearby excited states can increase the probability of off-resonant spontaneous emission, thus reducing the EIT efficiency \cite{Deng2001}, and can also distort the EIT window \cite{Chen2009}. In the case of alkali atoms this favours the use of the $D_{1}$ line for its simpler electronic structure.

\begin{figure}[!t]
\centering
\includegraphics[width=9.5cm]{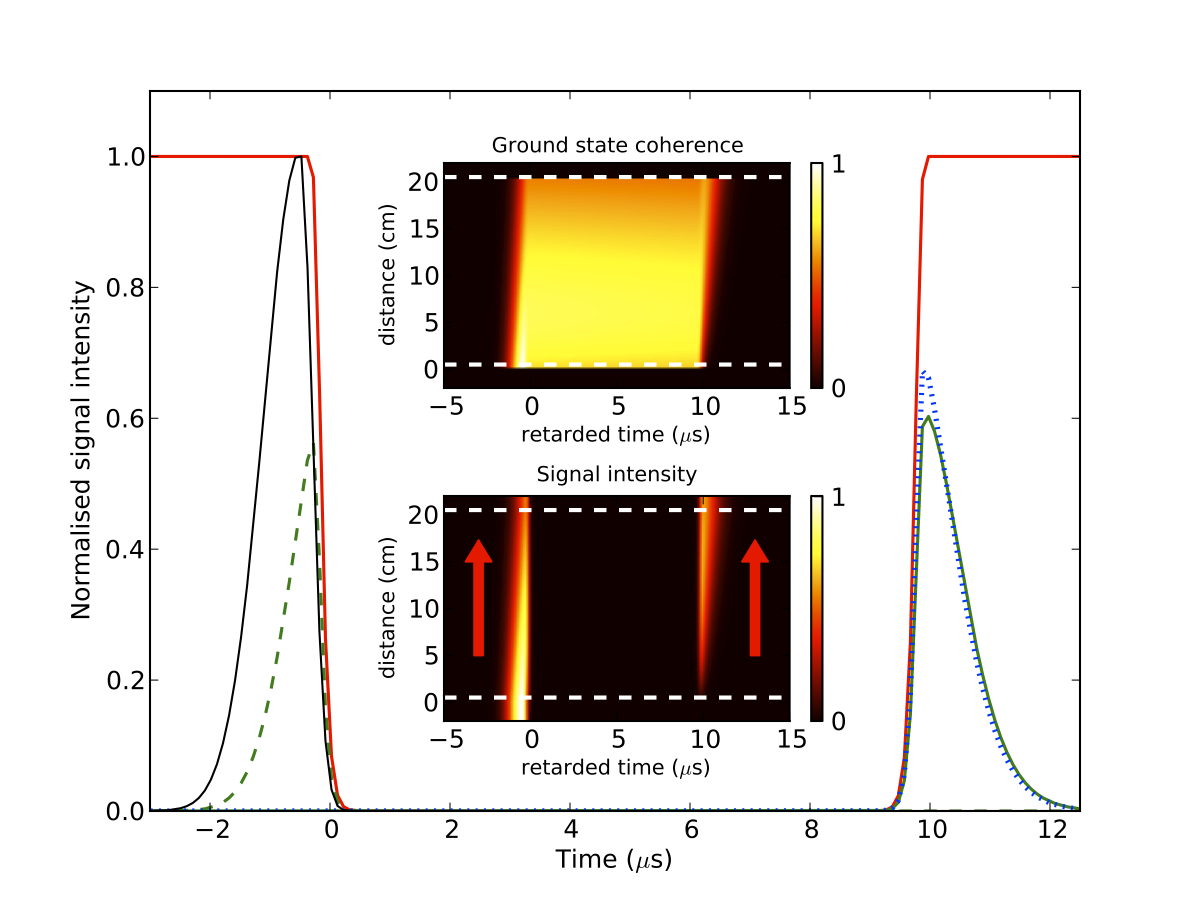}
\caption{Numerical solution of the density matrix equations for an photon pulse stored in a thermal vapour. The main plot shows the control beam (thick red solid line), the input signal field (black), and the signal output for co- (solid green line) and counter- (dotted blue line) propagating beams during retrieval. The green dashed lines shows the leakage of the signal beam before storage. We see that the EIT storage retains the shape of the pulse and that counterpropagating retrieval beams suffer less dispersion and loss. The reason for this is visible from the upper inset, which shows the spin wave (ground state coherence) stored mainly in the first section of the cell and therefore passes through less of the vapour cell during the reading stage. The lower insert shows the propagation of the signal field during the storage procedure.  The dashed white lines represent the entrance and exit positions of the vapour cell and in all plots the x-axis represents the retarded time $t_{r}=t-z/c$, where $z$ is the distance and $c$ is the vacuum velocity of light. The model assumes the vapour is heated to an optical depth of $d=$15 and includes Neon buffer gas with a collisional rate of $2\pi\,200\,$MHz.\label{LS1}}
\end{figure}

To verify that single photons with temporal length scales of $\sim1\,\mu$s can be stored in an EIT vapour cell we performed a numerical simulation to solve the density matrix equations for a simple open $\Lambda$ system. The transition strengths and linewidths are based upon the Zeeman sub levels in the $^{87}$Rb D$_{1}$ line. The control beam couples the $\langle F'=1,m_{F}'=0|F=1,m_{F}=-1\rangle $ transition and the signal beam (scaled to single photon intensity) couples the $\langle F'=1,m_{F}'=0|F=1,m_{F}=+1\rangle $ transition. The photons can be stored within a 20\,cm long vapour cell (as shown in Fig. \ref{LS1}) for co- and counter-propagating reading beams with a small amount of leakage using the boundary conditions from Eqn \ref{bounds}. As predicted earlier, the counterpropagating setup results in a more efficiency retrieval. The simulation also shows that the photon's temporal profile is retained with only simple control beam dynamics and the storage could be improved with optimization \cite{Gorshkov2007b}.

%---------------------------------------------------------------------------------------------------------------------------------
\section{Qubit Characteristics}
The suitability of photons produced using V-STIRAP for QIP has been demonstrated in several experiments. To seamlessly integrate EIT storage into a scalable quantum network we must be certain that the process does not interfere with the computationally important properties of the photons. Namely, the photons should retain the same degree of indistinguishability and entanglement after any EIT manipulations.
 
EIT has been shown to preserve entanglement between two short laser pulses by passing a linearly polarized pulse through a quarter waveplate and spatially separating the orthogonal polarizations with a beamsplitter thus creating an entangled pair. The two modes are then stored in a thermal EIT vapour, recombined and their correlations measured \cite{Choi2008}. We also note that recently storage of entangled photons has been demonstrated in a solid state device using a related quantum memory technique known as an atomic frequency comb \cite{Clausen2010}. 
 
There are several techniques to verify the statistical characteristics of photon sources. As mentioned earlier, the anti-bunching nature of a single photon source can be determined by the Hanbury-Brown-Twiss (HBT) technique, in which the photon stream is divided by a 50:50 beam-splitter and the temporal correlation between clicks from single photon counter modules (SPCM) at each output port are measured at different timescales \cite{Lounis2005}. A pure single photon source possesses no correlation at zero time delay.  
A measurement of the uniformity of the photon temporal profile can be made using a single photon interferometer described by Hong, Ou \& Mandel (HOM) \cite{Kuhn2010}. Two photons from the same source are combined at a 50:50 beam-splitter and the photon wavefunctions interfere such that if the photons arrive simultaneously with identical shapes, then the photons can only exit the beamsplitter through one port. Both anti-bunching \cite{Kuhn2002,McKeever2004,Keller2004} and HOM interferences \cite{Legero2004,Legero2006} have been demonstrated for V-STIRAP sources. 

The photons produced by an atom-cavity system tend to be on the order of $1\,\mu$s long, well over the time resolution of a SPCM ($\sim$ns). Therefore in addition to the standard HOM technique a quantum beat note of the emitted photons can be measured which can provide additional information regarding the coherence properties of the incident photons \cite{Legero2004}. A full characterisation of the photons may be gained by this technique as the temporal shape, detuning, and frequency jitter all give rise to different behaviours \cite{Legero2006}. By measuring the coherence properties of the photons from the cavity and EIT storage, it will be possible to define the length scales over which the quantum network may efficiently perform. 
 
\begin{figure}[!t]
\centering
\includegraphics[width=8.5cm]{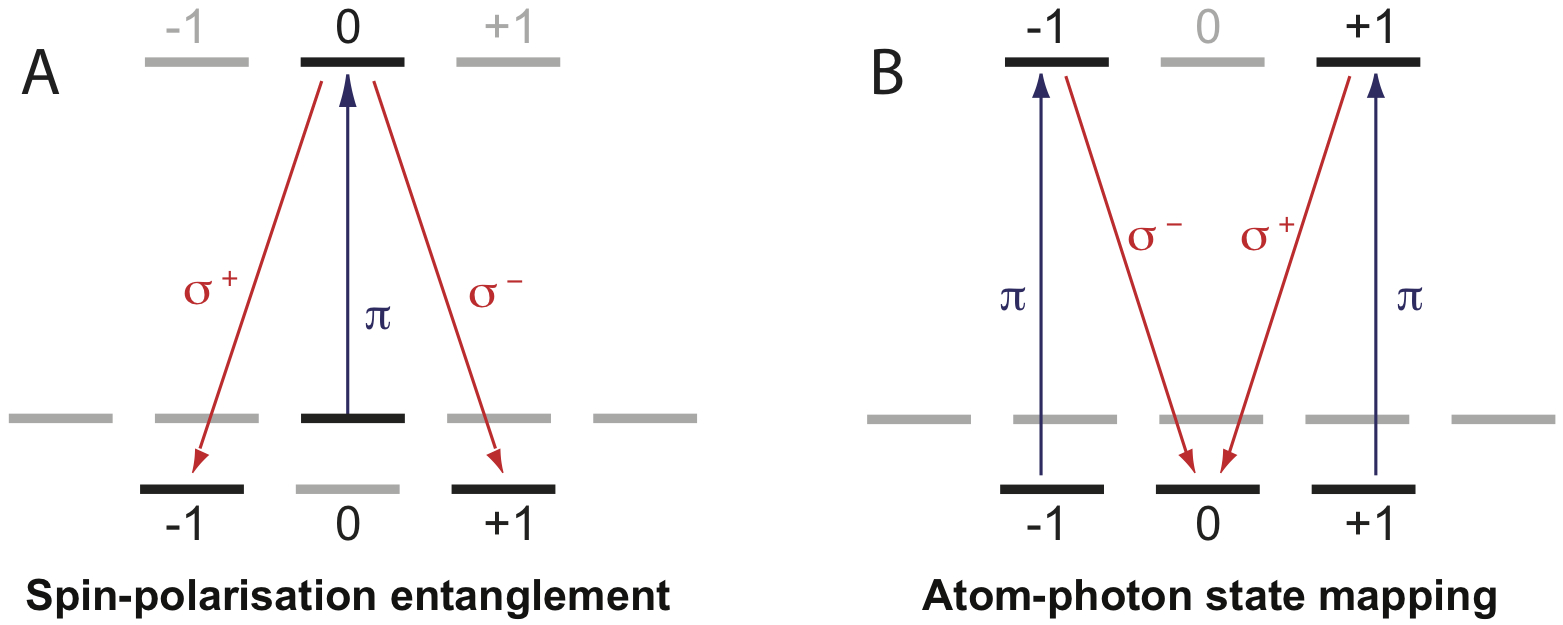}
\caption{Optical (blue) and cavity mode (red) structure in $^{87}$Rb to entangle consecutive photons emitted from the cavity \cite{Wilk2007}.  A) Atoms excited with a linearly ($\pi$) polarized drive laser may produce either $\sigma^{+}$ or $\sigma^{-}$ circularly polarized photons, thus leaving the first emitted photon and the atom in the entangled state $\psi_{A}=\frac{1}{\sqrt{2}} ( \vert -1,\sigma^{+}\rangle \pm \vert +1,\sigma^{-}\rangle)$. B) A second excitation with $\pi$ polarized light results in either  $\sigma^{+}$ or $\sigma^{-}$ photons which become entangled with the first photons with state $\psi_{B}=\frac{1}{\sqrt{2}} (\vert \sigma^{-},\sigma^{+}\rangle \pm \vert \sigma^{+},\sigma^{-}\rangle)$, and the atom is left disentangled. \label{entangle}}
\end{figure}

The preservation of entanglement between the cavity atom and the stored photon can be measured in an experiment demonstrated by Wilk \textit{et al} \cite{Wilk2007}. In their setup which involved the $^{87}$Rb D$_{2}$ line, an atom initially prepared in the $|F=2, m_{F}=0\rangle$ state is coupled by the `entangling' laser to $|F'=1, m'_{F}=0\rangle$ excited state by linearly polarized light (Fig. \ref{entangle}.A). The cavity couples the excited state to $|F=1\rangle$ and can support both $\sigma^{+}$ and  $\sigma^{-}$ modes which are emitted in the V-STIRAP process. This photon could then be stored in the EIT system (using the hyperfine splitting to maintain polarization independance) and a second photon is generated by the cavity by coupling the excited state to $F=1$ with linearly polarized beams (Fig. \ref{entangle} B). With the same cavity coupling, the second photon will have a polarization dependent upon the first photon and their correlation (and thus entanglement) can be measured through quantum state tomography \cite{Altepeter2005}. 

\section{Quantum Networking}
As mentioned in the introduction, a C-QED single photon source coupled with an EIT memory is an ideal candidate for repeat-until-success (RUS) quantum computing and we qualify this here. RUS is a form of measurement-based quantum computation which involves an array of stationary qubits that are not measured directly \cite{Lim2006}. Instead, ancillary qubits are measured in a mutually unbiased basis, so that an observer cannot gain knowledge about the stationary qubit's state. This provides a method of producing entanglement without the risk of destroying or decohering the qubits. An example is two C-QED photon sources (stationary qubits) each emitting a photon (ancillary qubits) which arrive onto a polarizing beamsplitter at the same time and are measured simultaneously in different output ports, so that the atoms within the cavities become entangled \cite{Duan2003}. This method can be used to construct graph states in which an array of qubits are individually prepared in a superposition state and connected (via the above method, for example) to form a maximally entangled resource. Also known as `cluster states', these can be used simulate any universal gate operation and therefore perform any quantum algorithm in a scaleable manner \cite{Raussendorf2001}. 

RUS provides a method to generate the cluster states via probabilistic, but heralded, two-qubit gates \cite{Lim2006}. We do not go into the details of the process here, but state that during the measurement stage of the qubit entanglement, a measurement of one or no photons results in a failure, and the measurement of a photon pair equals either a success or failure with the qubits remaining in their original state (due to the unbiased basis). This last result, known as \emph{failure with insurance}, can be converted to a sufficiently probable success by repeating the entanglement process. RUS quantum computing is robust to probabilistic entanglement operations because success is heralded, therefore it does not require near unity efficient photon sources (unlike LOQC) \cite{Lim2006}. The cluster size is limited by the decoherence time of the stationary qubits and the time to perform the graph state operations. The cluster can be extended during the QIP operation by providing entangled `buffer zones' \cite{Raussendorf2001}.  

\begin{figure}[!t]
\centering
\includegraphics[width=9.0cm]{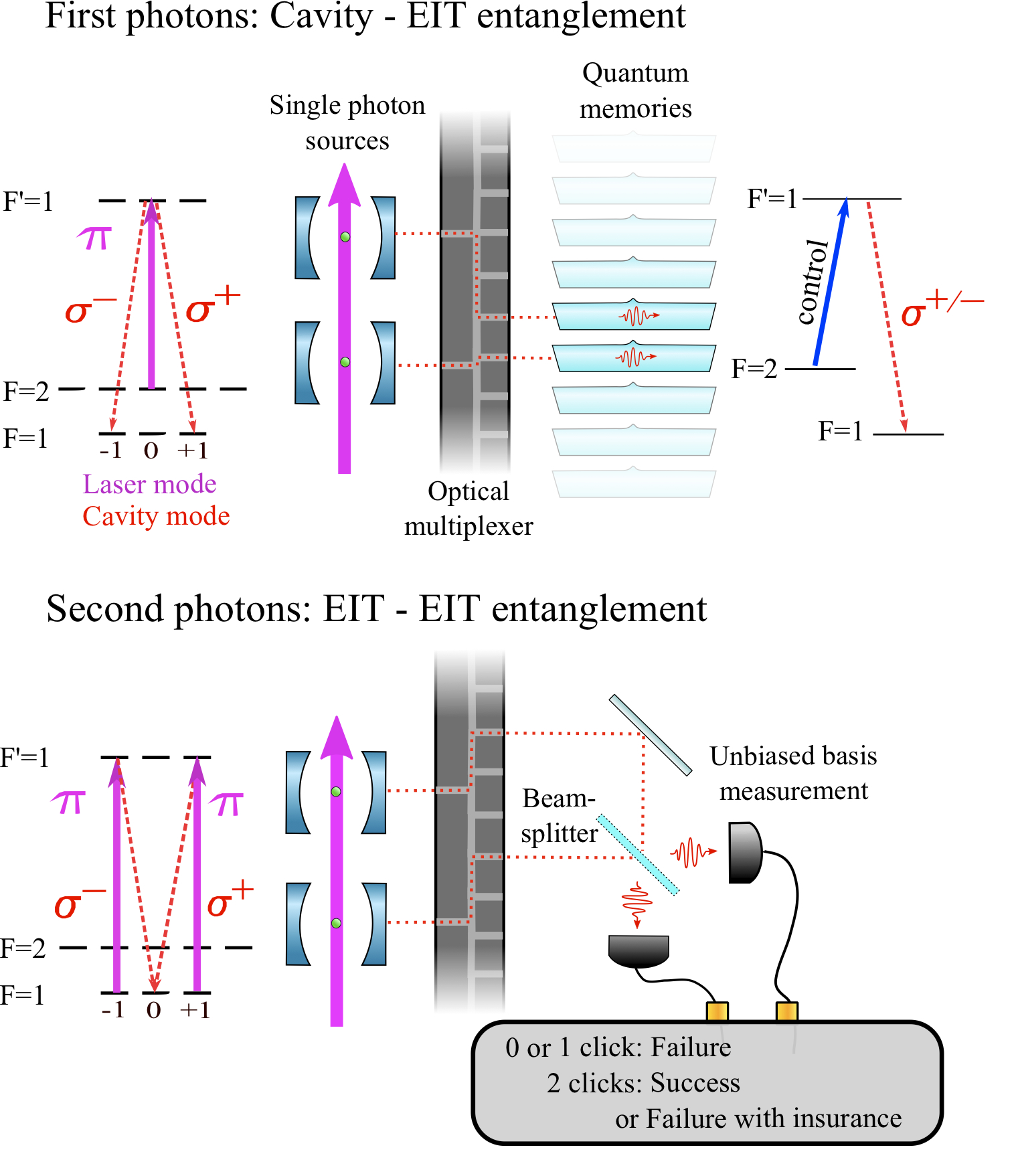}
\caption{Illustration of the stages of a repeat-until-success quantum computing scheme using quantum memories as a cluster state and a deterministic photon source to enable entanglement. See text for further details. The optical multiplexer would be a fast, polarization independent, optical switching array to route the photons between the sources and memories/detectors, for example see Ref \cite{Sugama2007}. \label{RUS}}
\end{figure}

The implementation of the RUS scheme presented by Lim \textit{et al.} \cite{Lim2006} consists of an array of cavity-based photon sources acting as stationary qubits. When using fixed emitters inside cavities (such as quantum dots, nitrogen vacancy centres, etc)  this scheme can indeed be scalable, but it may not be feasible using trapped or free atoms. This is due to the probabilistic nature of loading each atom into a cavity simultaneously, and becomes more difficult the larger the cluster. This problem may be overcome by using an array of EIT memory cells as stationary qubits and use only a small number of photon pistols. The scheme we propose is shown in Figure \ref{RUS} and proceeds as follows:
\begin{itemize}
\item
1. A small collection of C-QED photon sources are simultaneously set running.
\item
2. When two of the cavities begin to emit photons, the pumping scheme proposed in Ref. \cite{Wilk2007} (Fig. \ref{entangle}) is implemented, which results in the first emitted photons entangled with their `parent' atoms.
\item
3. The first photons are stored in an EIT cell, thus entangling the parent atoms to the EIT memory.
\item
4. The parent atoms then undergo the second photon emission which disentangles them, leaving the second photons entangled with the EIT memory.
\item
5. The second photons are then measured together in an unbiased basis which leaves the two EIT cells entangled.
\item
6. The two cavities can then be used to repeat the process with other EIT memories until the cluster state is formed. 
\end{itemize}        

The second photons are only emitted on condition of the first photon emission and therefore if the first photons are stored successfully then detection of the second photon assures entanglement between the EIT memories. This scheme also uses the long deterministic stream of photons more efficiently than the `standard' RUS scheme which stops the photon source once the entanglement success has been achieved. If any other cavities from the collection start to emit photons these can be used to entangle more EIT memories or take over from cavities which have lost their atoms. 

The limitation to this scheme is the decoherence time of the quantum memories which will reduce the fidelity of the cluster state algorithm. As mentioned earlier the entire cluster state does not have to be constructed before the algorithm is implemented, only those elements which are undergoing gate operations plus a buffer zone. As noted in \cite{Raussendorf2001}, this can be used so that the quantum memories do not suffer significant decoherence before they are used. We may also note that due to the use of deterministic single photon sources and quantum memories, this scheme can be directly linked with quantum repeater systems \cite{Dur1999} which provide a method of spatially extending quantum networks.

\section{Conclusion}
To summarize, the cavity V-STIRAP source can produce single photons deterministically with high probabilities with resilience to variations in cavity coupling. The photons are indistinguishable and may be tailored to specific shapes with dynamic control over the laser intensity. Their central frequency is defined by the optical resonance, and their bandwidth (and temporal length) is dependent upon the adiabatic passage between ground states. This is normally in the microsecond regime and is bounded by the interaction time of the atom within the cavity (ms) and by the cavity coupling rate ($\tau>1/g$, assuming $\Omega_{max}\gg g$). The repetition rate depends upon the preparation and pulse scheme and repitition rates upto a MHz have been demonstrated.

The photonic qubits can be stored for periods up to hundreds of microseconds in optically dense thermal vapours with high efficiency. Storage may be achieved on the transitions used for the single photon source, i.e. between hyperfine or Zeeman split substates. The limitations to the storage efficiency are nonlinear effects at high optical depths and spin wave dephasing due to inhomogeneous magnetic fields and diffusive motion of the atoms. The possibly detrimental effects of Doppler broadening may be overcome with the addition of buffer gas. There exist optimization procedures which result in an efficient and high fidelity re-mapping of the spin wave back into a photon. We have also discussed how the EIT storage should not affect the properties of the photons which are important to quantum information processing and we have outlined methods in which to characterize them.

Finally, We have presented a repeat-until-success quantum computation scheme using the above components which has potential for robust and efficient scalability.

%---------------------------------------------------------------------------------------------------------------------------------
\bibliographystyle{h-physrev3}
\bibliography{ref}

\end{document}